\documentclass[reprint,aip,numerical,amsmath,fleqn,groupedaddress]{revtex4-1}
\usepackage[mathlines]{lineno}
\modulolinenumbers[1]
\usepackage{graphicx}
\usepackage[hidelinks]{hyperref}

\begin{document}

\preprint{arXiv:1307.4511}

\title[Nuclear interactions in polyethylene range compensators]{Influence of nuclear interactions in polyethylene range compensators for carbon-ion radiotherapy}

\thanks{This study was in part presented at the 106th Scientific Meeting of the Japan Society of Medical Physics in Suita, Osaka, Japan,  16--18 September 2013 and at the 53rd Scientific Meeting of the Particle Therapy Co-Operative Group in Shanghai, China, 12--14 June 2014. It has also been fully published in Medical Physics 41(7) 071704, July 2014 (\url{http://dx.doi.org/10.1118/1.4870980}).}

\author{Nobuyuki Kanematsu}\email[]{nkanemat@nirs.go.jp}
\author{Yusuke Koba}
\author{Risa Ogata}
\affiliation{Research Center for Charged Particle Therapy, National Institute of Radiological Sciences, 4-9-1 Anagawa, Inage-ku, Chiba 263-8555, Japan}

\author{Takeshi Himukai}
\affiliation{Ion Beam Therapy Center, SAGA HIMAT Foundation, 415 Harakoga-machi, Tosu, Saga 841-0071, Japan}


\date{July 2013---June 2014, Reprint: \today}

\begin{abstract}
\begin{description}
\item[Purpose]
A recent study revealed that polyethylene (PE) would cause extra carbon-ion attenuation per range shift by 0.45\%/cm due to compositional differences in nuclear interactions.
The present study aims to assess the influence of PE range compensators on tumor dose in carbon-ion radiotherapy.
\item[Methods] 
Carbon-ion radiation was modeled to be composed of primary carbon ions and secondary particles, for each of which the dose and the relative biological effectiveness (RBE) were estimated at a tumor depth in the middle of spread-out Bragg peak.
Assuming exponential behavior for attenuation and yield of these components with depth, the PE effect on dose was calculated for clinical carbon-ion beams and was partly tested by experiment. 
The two-component model was integrated into a treatment-planning system and the PE effect was estimated in two clinical cases.
\item[Results]
The attenuation per range shift by PE was 0.1\%--0.3\%/cm in dose and 0.2\%--0.4\%/cm in RBE-weighted dose, depending on energy and range-modulation width.
This translates into reduction of RBE-weighted dose by up to 3\% in extreme cases.
In the treatment-planning study, however, the effect on RBE-weighted dose to tumor was typically within 1\% reduction.
\item[Conclusions]
The extra attenuation of primary carbon ions in PE was partly compensated by increased secondary particles for tumor dose.
In practical situations, the PE range compensators would normally cause only marginal errors as compared to intrinsic uncertainties in treatment planning, patient setup, beam delivery, and clinical response.
\end{description}
\end{abstract}


\maketitle

\section{Introduction}

The essence of radiotherapy with carbon-ion beams is conformation of a spread-out Bragg peak (SOBP), which is a delivered high-dose region concurrently enhanced by relative biological effectiveness (RBE), to a tumor.
In a conventional broad-beam system, a custom-shaped range compensator shortens extra beam range in the field to minimize exposure of normal tissue beyond the tumor.\cite{Urie-1983}
At the National Institute of Radiological Sciences (NIRS), the range compensators have variable water-equivalent (WE) thickness between 0.3 cm and 20 cm.
For the material, high-density polyethylene (PE) has been commonly used in carbon-ion radiotherapy facilities.\cite{Kanai-1999, Hishikawa-2004, Ohno-2011} 

It was revealed recently that range shifting with PE would cause extra attenuation of carbon ions by 0.45\%/cm as compared to water,\cite{Kanematsu-2013} due to non water equivalence in nuclear interactions originated from compositional differences. 
In dosimetry, dose to water is often measured in non-water phantom with correction by fluence correction factor, which is a predetermined ratio of dose-in-water to dose-to-water in the phantom.
The fluence correction factor varies among beams, materials, and depths and can be specifically determined by experiment or Monte Carlo simulation.\cite{Palmans-2002, Schneider-2002, Al-Sulaiti-2010, Luhr-2011, Moyers-2011, Al-Sulaiti-2012, Palmans-2013, Rossomme-2013} 

The non water equivalence in nuclear interactions has been long ignored in treatment planning, beam delivery, and clinical studies.
It is therefore urgent to assess realistic influence of the PE range compensators on carbon-ion radiotherapy.
The PE effect on dose can be evaluated in the same manner as for the fluence correction factor, which would, however, be impractical because clinical carbon-ion beams are specific to facilities with a wide variety of energies and range modulations.
In addition, RBE of the carbon-ion beams varies with depth in water. 
These complexities would naturally demand radical simplification in physical and biological modeling of the PE effect to enable clinical dose evaluation. 

\section{Materials and methods}

\subsection{Modeling framework}\label{sec:2A}

\subsubsection{The treatment system}

For conventional carbon-ion radiotherapy at NIRS except for rare ocular-melanoma cases, carbon ions are accelerated to an energy per nucleon of $E/A = 290$, 350, 400, or 430 MeV by the Heavy Ion Medical Accelerator in Chiba (HIMAC).
The beam is extracted and transported to one of three treatment rooms, where it is broadened three-dimensionally to form a clinical carbon-ion beam as follows:\cite{Torikoshi-2007}
A beam-wobbling system, which is composed of a pair of electromagnets and a variable scatterer, wobbles the scattered beam in a circular orbit to form a uniform field of 10, 15, or 20 cm in diameter.\cite{Yonai-2008b}
One of 13 exchangeable ridge filters modulates the beam range to a designed width between 2 cm and 15 cm in WE length and forms a SOBP in the depth--dose distribution.
A range shifter finely degrades the energy and shortens the range from those specific to the four extraction energies.

Scattering variation induced by the range modulation and shifting necessitates intermittent adjustment of the wobbling condition with range shift to retain sufficient field uniformity.
For all these combinations, there are currently about 500 clinical carbon-ion beams in the treatment-planning system (XiO-N, Mitsubishi Electric Corporation, Tokyo) registered with associated beam data including relative dose $D_w$ and RBE $\epsilon_w$ as a function of depth $d$ in water. While the dose data were measured, the RBE data were originated from ridge-filter design by Monte Carlo simulation integrated with radiobiology models to address mixing of carbon ions and a wide spectrum of secondary particles.\cite{Kanai-1999, Kase-2006, Sakama-2012}

In treatment planning, among the clinical carbon-ion beams, the one that minimally covers a given tumor target with its SOBP is selected and conformed to the target with a collimator and a range compensator.\cite{Kanematsu-2007}
The monitor-unit number is experimentally determined for each treatment beam to deliver the prescribed dose at a reference depth in water without the range compensator.
As every target tumor should be in the SOBP, the reference depth is set to the mid-SOBP depth:
\begin{eqnarray}
d_\mathrm{mid} = R - \frac{M}{2},
\end{eqnarray}
where $R$ is the unmodulated beam range and $M$ is the range-modulation width.
Because depth-dependent analysis may be beyond this study intended for assessment of clinical influence, we simply evaluate dose and RBE in the middle of SOBP: mid-SOBP dose $D_\mathrm{mid} = D_w(d_\mathrm{mid})$ and mid-SOBP RBE $\epsilon_\mathrm{mid} = \epsilon_w(d_\mathrm{mid})$, unless noted otherwise.

\subsubsection{Nuclear interactions}

In conventional analytic modeling of nuclear interactions in proton beams,\cite{Bortfeld-1997, Palmans-2013} the secondary particles were assumed to deposit dose locally, which may be appropriate for target-nucleus recoil and fragments produced in nuclear breakup at rest.
In carbon-ion beams, the dose originated from target nuclei is generally trivial in comparison to high ionization of carbon ions.
Instead, light nuclei constantly produced in matter by fragmentation of projectile carbon ions dominate the secondary-particle dose and penetrate far beyond the carbon ions.\cite{Matsufuji-2003}
We assumed that neither the target nuclide nor incident energy strongly influences how the projectile nucleus may break up,\cite{Goldhaber-1974} which was experimentally valid for carbon-ion beams.\cite{Matsufuji-2005}
Although about as many neutrons as protons may be produced via nuclear fragmentation, the neutrons mostly escape from the field and are only relevant to the stochastic effect.\cite{Yonai-2008a}
 
\subsubsection{Two-component model}

We model a carbon-ion radiation as a mixture of primary carbon ions and secondary particles, in which the mid-SOBP dose is decomposed into respective contributions:
\begin{eqnarray}
D_ \mathrm{mid} = D_1 + D_2.
\end{eqnarray}
In the exponential-attenuation approximation, carbon-ion dose $D_1$ will decrease with PE range shift $s$ by the relative carbon-ion attenuation factor: 
\begin{eqnarray}
f_1 = e^{-\kappa s} \approx 1-\kappa\, s,
\end{eqnarray}
where $\kappa = 0.45\%/\mathrm{cm}$ is the relative carbon-ion attenuation coefficient for PE range shift.\cite{Kanematsu-2013}
The secondary particles will increase inversely with reduction of the primary carbon ions.
In the first-order approximation, secondary-particle dose $D_2$ will thus increase with PE range shift $s$ by the relative secondary-particle yield factor:
\begin{eqnarray}
f_2 \approx \frac{1}{f_1} = e^{\kappa s} \approx 1+\kappa\, s.
\end{eqnarray}

\subsubsection{Dose modification}

We define dose modifying factor $f_D$ as the ratio of dose $D_{\mathrm{mid},s}$ with PE range shift $s$ to dose $D_\mathrm{mid}$ without PE, which is the inverse of the fluence correction factor for dosimetry. 
In the two-component model, it is formulated as
\begin{eqnarray}
f_D = \frac{D_{\mathrm{mid},s}}{D_\mathrm{mid}} = \frac{f_1 D_1 + f_2 D_2}{D_1 + D_2} \approx 1-\kappa_D \, s, 
\end{eqnarray}
where $\kappa_D$ is the dose attenuation coefficient given by
\begin{eqnarray}
\kappa_D = \frac{\frac{D_1}{D_2}-1}{\frac{D_1}{D_2}+1}\, \kappa.
\end{eqnarray}

\subsubsection{RBE modifification}

The RBE that has been used for carbon-ion radiotherapy at NIRS was originally defined for an in vitro endpoint: survival fraction $S = 10\%$ for human salivary gland (HSG) cells,\cite{Furusawa-2000} as
\begin{eqnarray}
\epsilon = \frac{2 \beta D_\gamma}{\sqrt{\alpha^2-4\beta\ln S}-\alpha},
\end{eqnarray}
where $\alpha$ and $\beta$ are the linear and quadratic dose coefficients of the HSG-cell survival response for the radiation of interest and $D_\gamma$ is the reference photon dose for the same endpoint.

In the lesion-additivity model,\cite{Lam-1987, Tilly-1999} the RBE of the two-component radiation is given by a dose-weighted average:
\begin{eqnarray}
\epsilon_\mathrm{mid} = \frac{\epsilon_1 D_1+ \epsilon_2 D_2}{D_1+D_2},
\end{eqnarray}
where $\epsilon_1$ and $\epsilon_2$ are the RBEs for the primary carbon ions and for the secondary particles.

We define RBE modifying factor $f_\epsilon$ as the ratio of the RBE with PE range shift to that without PE:
\begin{eqnarray}
f_\epsilon = \left(\frac{\epsilon_1 f_1 D_1 + \epsilon_2 f_2 D_2}{f_1 D_1+f_2 D_2}\right)
\left(\frac{\epsilon_1 D_1+ \epsilon_2 D_2}{D_1+D_2}\right)^{-1}.
\end{eqnarray}

\subsubsection{Clinical-dose modification}

In carbon-ion radiotherapy at NIRS, clinical doses are prescribed in RBE-weighted dose with extra scaling typically by factor 1.44 to accommodate to historical treatment protocols.\cite{Matsufuji-2007}
The clinical dose per monitor unit will vary with PE range shift $s$ by a clinical-dose modifying factor:
\begin{equation}
f_{\epsilon D} = f_\epsilon \, f_D 
= \frac{\epsilon_1 f_1 D_1 + \epsilon_2 f_2 D_2}{\epsilon_1 D_1+ \epsilon_2 D_2}
\approx 1-\kappa_{\epsilon D}\, s,
\end{equation} 
where $\kappa_{\epsilon D}$ is the clinical-dose attenuation coefficient given by
\begin{eqnarray}
\kappa_{\epsilon D} = \frac{\frac{\epsilon_1}{\epsilon_2} \frac{D_1}{D_2} - 1}{\frac{\epsilon_1}{\epsilon_2} \frac{D_1}{D_2} + 1}\, \kappa. 
\end{eqnarray}

\subsubsection{Energy dependence} 

Parameters $D_1/D_2$ and $\epsilon_1/\epsilon_2$ are specific to individual ridge filters. 
In a group of clinical carbon-ion beams sharing the same ridge filter, $D_1/D_2$ will vary with extraction energy or beam range due to attenuation of the primary carbon ions and yield of the secondary particles by nuclear interactions while $\epsilon_1/\epsilon_2$ may be approximately invariant.
Dose $D_1$ may decrease with the primary carbon ions that attenuate exponentially with attenuation length $\lambda_w = 25.5$ cm in water,\cite{Kanematsu-2013} and $D_2$ may increase just inversely in the first-order approximation.
For a beam of range $R$, the dose ratio may thus be given by
\begin{eqnarray}
\frac{D_1}{D_2} \approx 
e^{2 \frac{R_0-R}{\lambda_w}}
\left(\frac{D_1}{D_2}\right)_{\!\!0}, 
\end{eqnarray}
where $R_0$ and $(D_1/D_2)_0$ are the beam range and dose ratio of the ridge filter for reference energy $E_0/A = 350$ MeV. 
Using an approximate range--energy relation,\cite{Kanematsu-2008} the beam range was estimated to be 16.3 cm for 290 MeV, 22.2 cm for 350 MeV, 27.6 cm for 400 MeV, and 31.0 cm for 430 MeV, where we ignored the energy loss mainly in the variable scatterer.
The energy factors for $D_1/D_2$ relative to 350 MeV then resulted in $e^{2(R_0-R)/\lambda_w} = 1.59$ for 290 MeV, 0.66 for 400 MeV, and 0.50 for 430 MeV.

\subsubsection{Ridge-filter design data}

\begin{figure}
\includegraphics[width=\columnwidth]{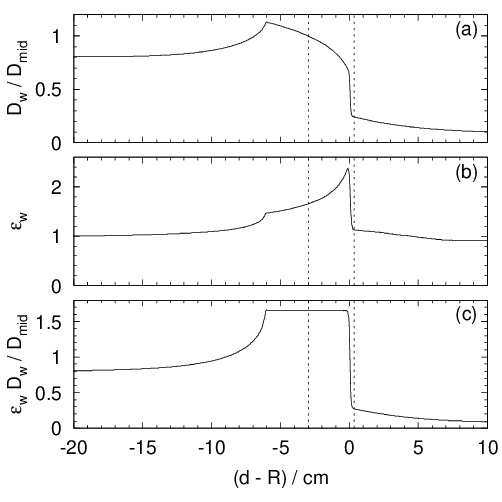}
\caption{\label{fig:1} Ridge-filter design data for extraction energy $E/A = 350$ MeV and range-modulation width $M = 6$ cm: (a) relative dose $D_w / D_\mathrm{mid}$, (b) RBE $\epsilon_w$, and (c) RBE-weighted dose $\epsilon_w D_w / D_\mathrm{mid}$, as a function of range-subtracted depth in water, $d-R$. Vertical dotted lines indicate the mid-SOBP and tail-base depths.}
\end{figure}

The ridge filters had been designed to modulate the carbon-ion range to deliver uniform RBE-weighted dose over the SOBP region for reference energy $E_0/A = 350$ MeV. 
Figure \ref{fig:1} shows the depth--dose and depth--RBE distributions designed for a ridge filter of 6-cm range modulation, for example.
We assumed that the quantity and composition of the secondary particles would not radically vary with depth up to the tail region, where primary carbon ions are absent.
This leads to approximations: $D_2 \approx D_\mathrm{tail} = D_w(d_\mathrm{tail})$ and $\epsilon_2 \approx \epsilon_\mathrm{tail} = \epsilon_w(d_\mathrm{tail})$, where we chose tail-base depth $d_\mathrm{tail}$ to be at 3 mm deeper than the carbon-ion range.
The dose and RBE ratios at the mid-SOBP depth were then estimated by
\begin{eqnarray}
\frac{D_1}{D_2} \approx \frac{D_\mathrm{mid}-D_\mathrm{tail}}{D_\mathrm{tail}}
\end{eqnarray}
and
\begin{eqnarray}
\frac{\epsilon_1}{\epsilon_2} \approx \frac{\epsilon_\mathrm{mid}}{\epsilon_\mathrm{tail}} + \left( \frac{\epsilon_\mathrm{mid}}{\epsilon_\mathrm{tail}} - 1 \right) \frac{D_\mathrm{tail}}{D_\mathrm{mid}-D_\mathrm{tail}},
\end{eqnarray}
as shown in Table \ref{tab:1}.
The $\epsilon_\mathrm{tail}$ values were reasonably invariant among the ridge filters. 
The apparent $D_\mathrm{tail}/D_\mathrm{mid}$ invariance among the ridge filters was due to incidental cancellation between $D_\mathrm{tail}$, which varied with modulation $M$, and $D_\mathrm{mid} = D_w(d_\mathrm{mid})$, which varied with proximal shift of mid-SOBP depth $d_\mathrm{mid} = R-M/2$.

\begin{table}
\caption{\label{tab:1} Ridge-filter design parameters for reference energy $E_0/A = 350$ MeV: ridge-filter identifier RF, range modulation $M$, relative tail-base dose $D_\mathrm{tail}/D_\mathrm{mid}$, tail-base RBE $\epsilon_\mathrm{tail}$, mid-SOBP RBE $\epsilon_\mathrm{mid}$, primary-to-secondary dose ratio $(D_1/D_2)_0$, and RBE ratio $\epsilon_1/\epsilon_2$.}
\setlength{\tabcolsep}{4pt}
\begin{tabular}{ccccccc}
\hline\hline
RF & $\frac{M}{\mathrm{cm}}$ & $\frac{D_\mathrm{tail}}{D_\mathrm{mid}}$ & $\epsilon_\mathrm{tail}$ & $\epsilon_\mathrm{mid}$ & $\left(\frac{D_1}{D_2}\right)_{\!\!0}$ & $\frac{\epsilon_1}{\epsilon_2}$ \\ 
\hline
020 & ~2.0 & 0.221 & 1.160 & 2.083 & 3.52 & 2.02 \\
025 & ~2.5 & 0.229 & 1.156 & 1.993 & 3.37 & 1.94 \\
030 & ~3.0 & 0.233 & 1.153 & 1.917 & 3.29 & 1.86 \\
040 & ~4.0 & 0.239 & 1.146 & 1.806 & 3.18 & 1.76 \\
050 & ~5.0 & 0.242 & 1.138 & 1.722 & 3.13 & 1.68 \\
060 & ~6.0 & 0.242 & 1.130 & 1.660 & 3.13 & 1.62 \\
070 & ~7.0 & 0.241 & 1.124 & 1.604 & 3.15 & 1.56 \\
080 & ~8.0 & 0.239 & 1.118 & 1.563 & 3.18 & 1.52 \\
090 & ~9.0 & 0.238 & 1.113 & 1.521 & 3.20 & 1.48 \\
100 & 10.0 & 0.235 & 1.110 & 1.486 & 3.26 & 1.44 \\
110 & 11.0 & 0.233 & 1.108 & 1.465 & 3.29 & 1.42 \\
120 & 12.0 & 0.230 & 1.107 & 1.431 & 3.35 & 1.38 \\
150 & 15.0 & 0.224 & 1.106 & 1.361 & 3.46 & 1.30 \\
\hline\hline
\end{tabular}
\end{table}

\subsection{Evaluation of modifying factors}

\subsubsection{Beam experiment}

We conducted an experiment with the apparatus identical to the previous study.\cite{Kanematsu-2013}
We tested some of the clinical carbon-ion beams, in which 430-MeV beams were excluded as they have been only rarely used. 
The central-axis doses for fields 10 cm in diameter were measured with special interest in the mid-to-distal part of the SOBP, with and without insertion of a 10-cm PE block in front of a dosimetric water phantom.
When the block was in, the phantom was moved downstream by 10 cm so that the SOBP would stay in the same position in the laboratory frame of reference for invariant beam-divergence effect.
For each tested beam, we determined the mid-SOBP doses and beam-range depths with and without PE insertion by fitting to the dose distribution designed for the ridge filter, which led to dose modifying factor $f_D$, range shift $s$, and dose attenuation coefficient $\kappa_D = (1-f_D)/s$.

\subsubsection{Model calculation}
 
While the dose modifying factor could be experimentally measured, it would not represent the clinical PE effect on RBE-weighted dose.
We thus calculated the dose and clinical-dose modifying factors for every combination of energy and modulation to cover all the clinical carbon-ion beams, according to the formulation described in Sec.~\ref{sec:2A}.

\subsection{Treatment planning}

\subsubsection{Model integration}

The treatment-planning system uses the pencil-beam algorithm, with which the clinical-dose distribution is calculated by two-dimensional convolution in the $x$--$y$ plane perpendicular to the field-central $z$ axis,
\begin{eqnarray}
\epsilon D(x,y,z) &=& \int\!\!\!\!\int \Phi_{x' y' z} K_{x' y' z}(x,y,z) dx' dy' ,
\end{eqnarray}
where $\Phi_{x' y' z}$ and $K_{x' y' z}$ are the in-air fluence and the convolution kernel on an axis directed to point $(x', y', z)$.\cite{Petti-1992}
The in-air fluence distribution was analytically modeled for the wobbling system.\cite{Tomura-1998, Kanematsu-2006} 
The kernel, which is a dose distribution for a virtual pencil beam, is defined as
\begin{eqnarray}
K_{x' y' z}(x,y,z) = \frac{\epsilon_w D_w(d_{x' y' z})}{2 \pi \sigma_{x' y' z}^2}\, e^{-\frac{(x-x')^2+(y-y')^2}{2 \sigma_{x' y' z}^2}},
\end{eqnarray}
where $\epsilon_w D_w$ is the in-water RBE-weighted dose per in-air fluence and $d_{x' y' z}$ and $\sigma_{x' y' z}$ are the WE depth and the rms off-axial spread at point $(x', y', z)$.\cite{Kanematsu-2008, Kanematsu-2009}

For this study, we tentatively modified the dose kernel to reflect the PE effect represented at the mid-SOBP depth: 
\begin{eqnarray}
K_{x' y' z} \rightarrow f_{\epsilon D}(s)\, K_{x' y' z},
\end{eqnarray}
where $f_{\epsilon D}$ is the clinical-dose modifying factor as a function of WE thickness $s$ of the compensator on the kernel axis.

\subsubsection{Case study}

To investigate the PE effect on tumor dose, we made a treatment-planning study.
We took two cases of typical indications for carbon-ion radiotherapy: chordoma of sacrum and malignant melanoma of nasal cavity and paranasal sinus.
As these tumors tend to be large and shallow, oblique-incident beams with large range compensation may often be applied.

For the sacrum case, we used real planning CT images of a prone patient, in which a clinical target volume had been delineated by a radiation oncologist.
In this simulation study, we did not apply special techniques or optimization processes normally used in clinical practice but simply planned one vertical and two opposing horizontal beams to treat the target. 
Similarly, for the nose-and-sinus case, we used real planning CT images of a supine patient, for which we simply planned one vertical and one horizontal beams.
Range compensators were designed to compensate the variation of target exit depth in the individual fields. 

The original and modified dose kernels were used to calculate plan dose distributions. 
For each beam in the original-kernel plan, the clinical dose of 1 Gy (RBE) was delivered to the center of the target placed at the machine isocenter.
For the beams in the modified-kernel plan, the same monitor-unit numbers as for the corresponding beams in the original-kernel plan were set.

\section{Results}

\subsection{Beam experiment}

\begin{figure}
\includegraphics[width=\columnwidth]{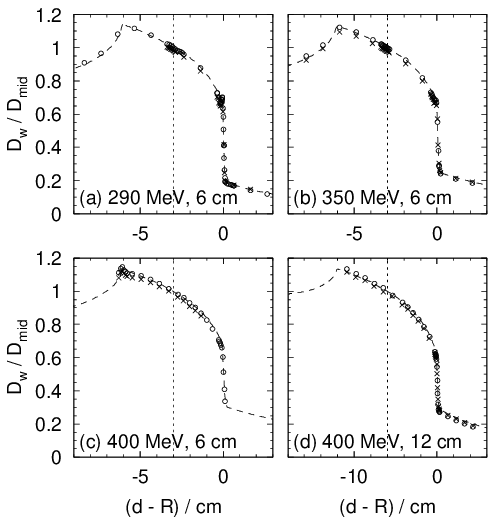}
\caption{\label{fig:2} Depth--dose distributions for some clinical carbon-ion beams: with 10-cm PE ($\times$), without PE ($\bigcirc$), and design (dashed line). 
The vertical dotted lines indicate mid-SOBP depths in range-subtracted depth $(d-R)$ and the texts indicate extraction energies and range-modulation widths.}
\end{figure}

Figure \ref{fig:2} shows examples of the depth--dose distributions measured in the experiment. 
Although the 10-cm PE should have caused extra carbon-ion attenuation by 4.5\%, the dose reduction was within 3\% as shown in Table \ref{tab:2}.
The relative dose reproducibility was typically $\pm 0.2\%$, which led to 0.3\% for the uncertainty of $f_D$.

\begin{table}
\caption{\label{tab:2} Measurement of dose modifying factors for clinical carbon-ion beams: beam identifier \#, energy $E/A$, ridge-filter identifier RF, range-modulation width $M$, range shift $s$, dose modifying factor $f_D$, and corresponding figure.}
\setlength{\tabcolsep}{3pt}
\begin{tabular}{ccccccc}
\hline\hline
\# & $\frac{E/A}{\mathrm{MeV}}$ & RF & $\frac{M}{\mathrm{cm}}$ & $\frac{s}{\mathrm{cm}}$ & $f_D$ & Fig. \\
\hline
~1 & 290 & 020 & ~2.0 & ~9.99 & 0.974 \\
~2 & 290 & 040 & ~4.0 & ~9.99 & 0.979 \\
~3 & 290 & 060 & ~6.0 & 10.00 & 0.982 & \ref{fig:2}(a) \\
~4 & 350 & 040 & ~4.0 & ~9.98 & 0.975 \\
~5 & 350 & 060 & ~6.0 & ~9.99 & 0.979 & \ref{fig:2}(b) \\
~6 & 350 & 080 & ~8.0 & ~9.99 & 0.979 \\
~7 & 400 & 040 & ~4.0 & ~9.99 & 0.979 \\
~8 & 400 & 060 & ~6.0 & ~9.99 & 0.985 & \ref{fig:2}(c) \\
~9 & 400 & 080 & ~8.0 & ~9.98 & 0.983 \\
10 & 400 & 120 & 12.0 & ~9.98 & 0.985 & \ref{fig:2}(d) \\
\hline\hline
\end{tabular}
\end{table}

\subsection{Model calculation}

\begin{figure}
\includegraphics[width=\columnwidth]{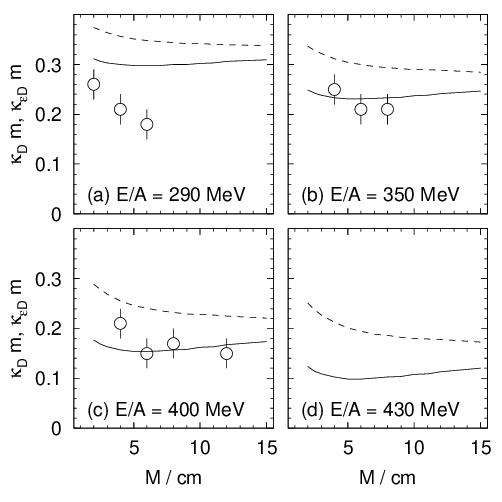}
\caption{\label{fig:3} Dose and clinical-dose attenuation coefficients $\kappa_D$ and $\kappa_{\epsilon D}$ in units of m$^{-1}$ or \%/cm, as a function of range-modulation width $M$: solid lines for $\kappa_D$ calculations, dashed lines for $\kappa_{\epsilon D}$ calculations, and open-circle symbols for $\kappa_D$ measurements.}
\end{figure}

Figure \ref{fig:3} shows the dose and clinical-dose attenuation coefficients for PE range shift as a function of range-modulation width, where we assumed functional continuity for the consistent design among the ridge filters.
They amounted to 0.1\%--0.3\%/cm for $f_D$ and 0.2\%--0.4\%/cm for $f_{\epsilon D}$, which were smaller than the carbon-ion attenuation coefficient of 0.45\%/cm due to partial compensation by increased secondary particles. 
Their systematic decrease with extraction energy was caused by the secondary particles increasing with beam range.
While the calculated and measured dose attenuation coefficients were consistent with each other for 350 MeV and 400 MeV, there was significant disagreement for 290 MeV.

\subsection{Case study}

\begin{figure}
\includegraphics[width=\columnwidth]{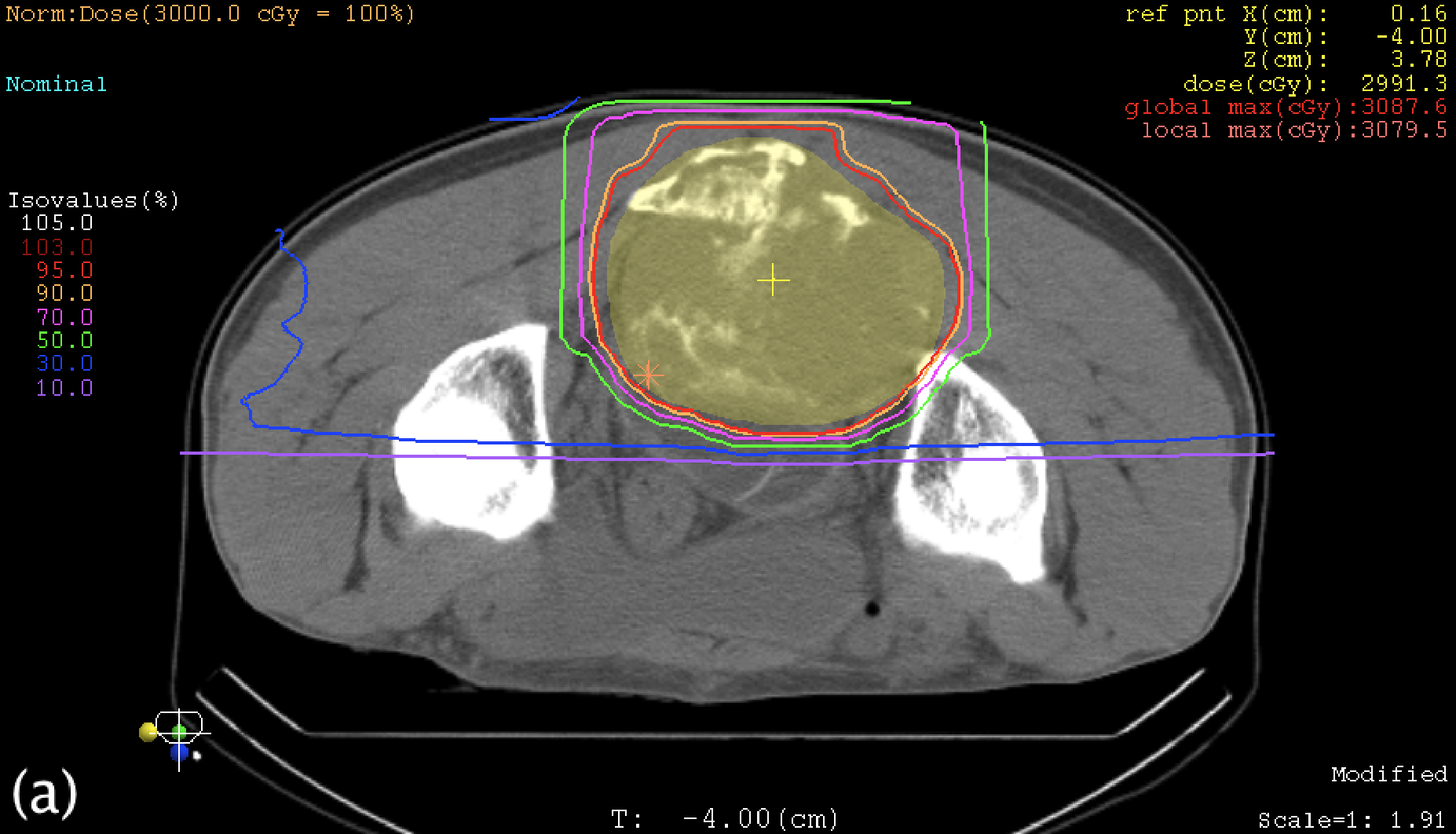}\\
\includegraphics[width=\columnwidth]{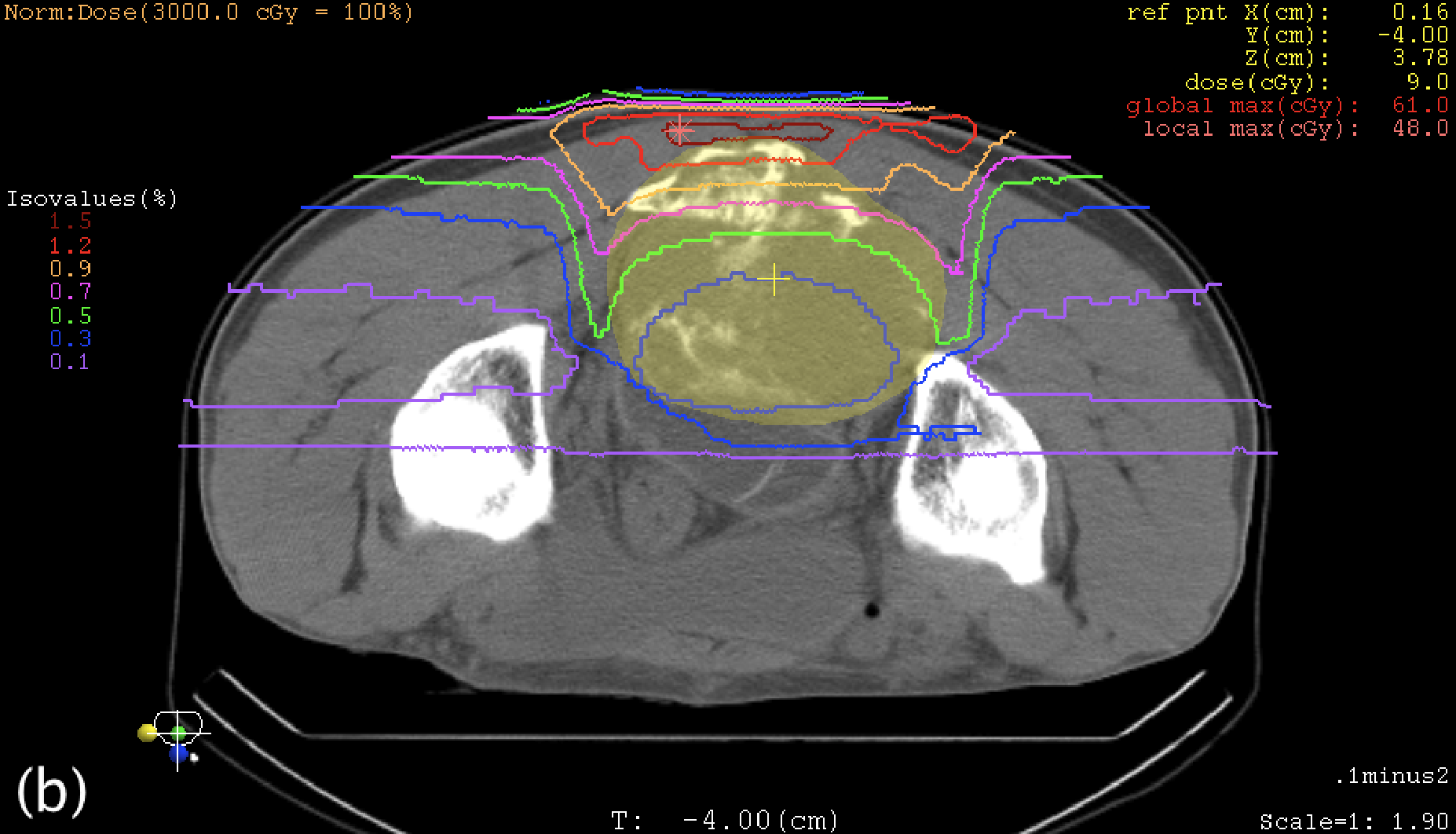}
\caption{\label{fig:4} Clinical-dose distribution on the isocenter ($+$) plane of trans-axial CT image for the sacrum target (yellowish region): (a) the modified-kernel plan with 95\%--10\% (red--purple) isodose contours and (b) dose reduction from the original-kernel plan with 1.5\%--0.1\% (dark red--purple) isodose contours.}
\end{figure}

\begin{figure}
\includegraphics[width=\columnwidth]{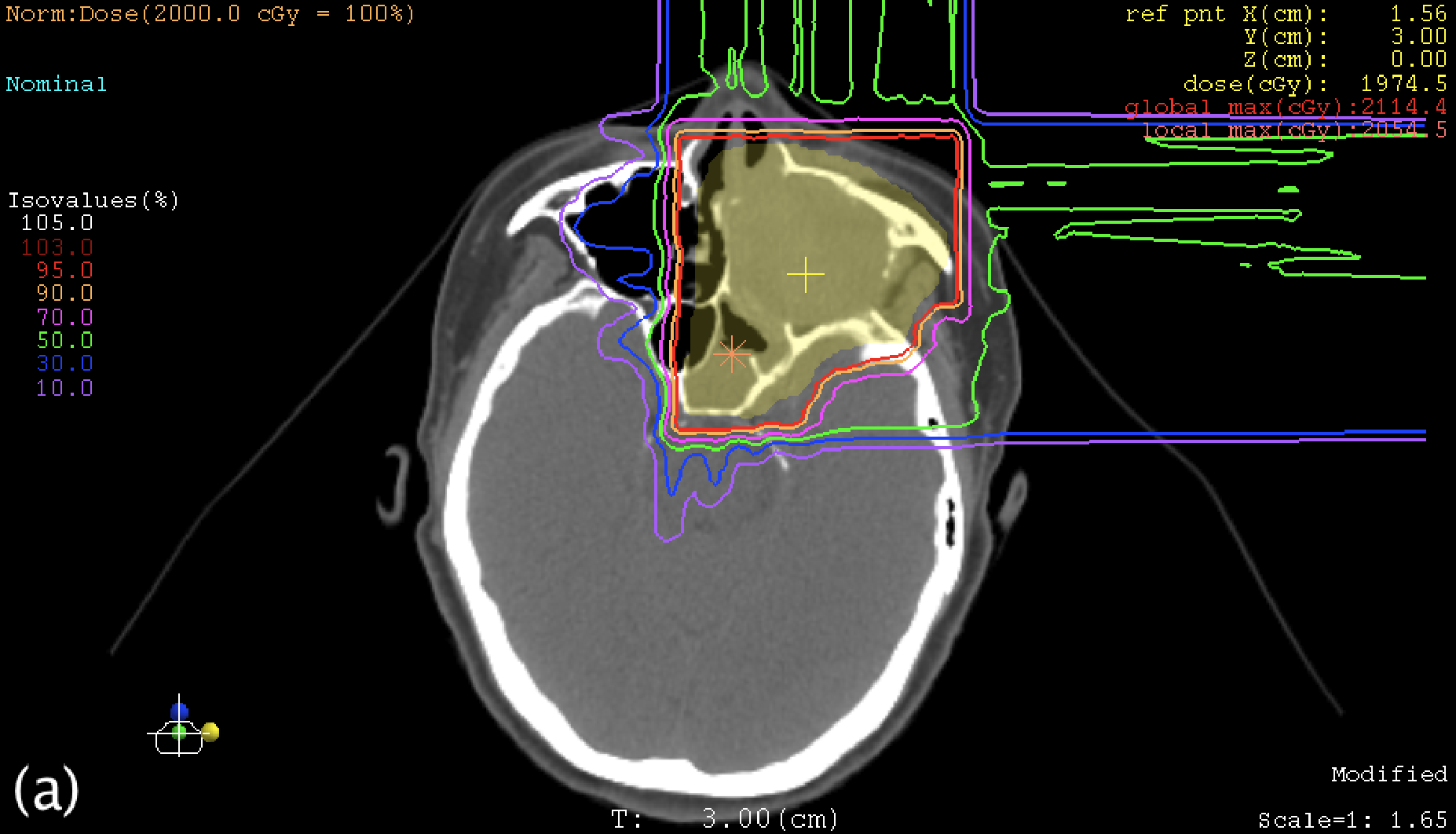}\\
\includegraphics[width=\columnwidth]{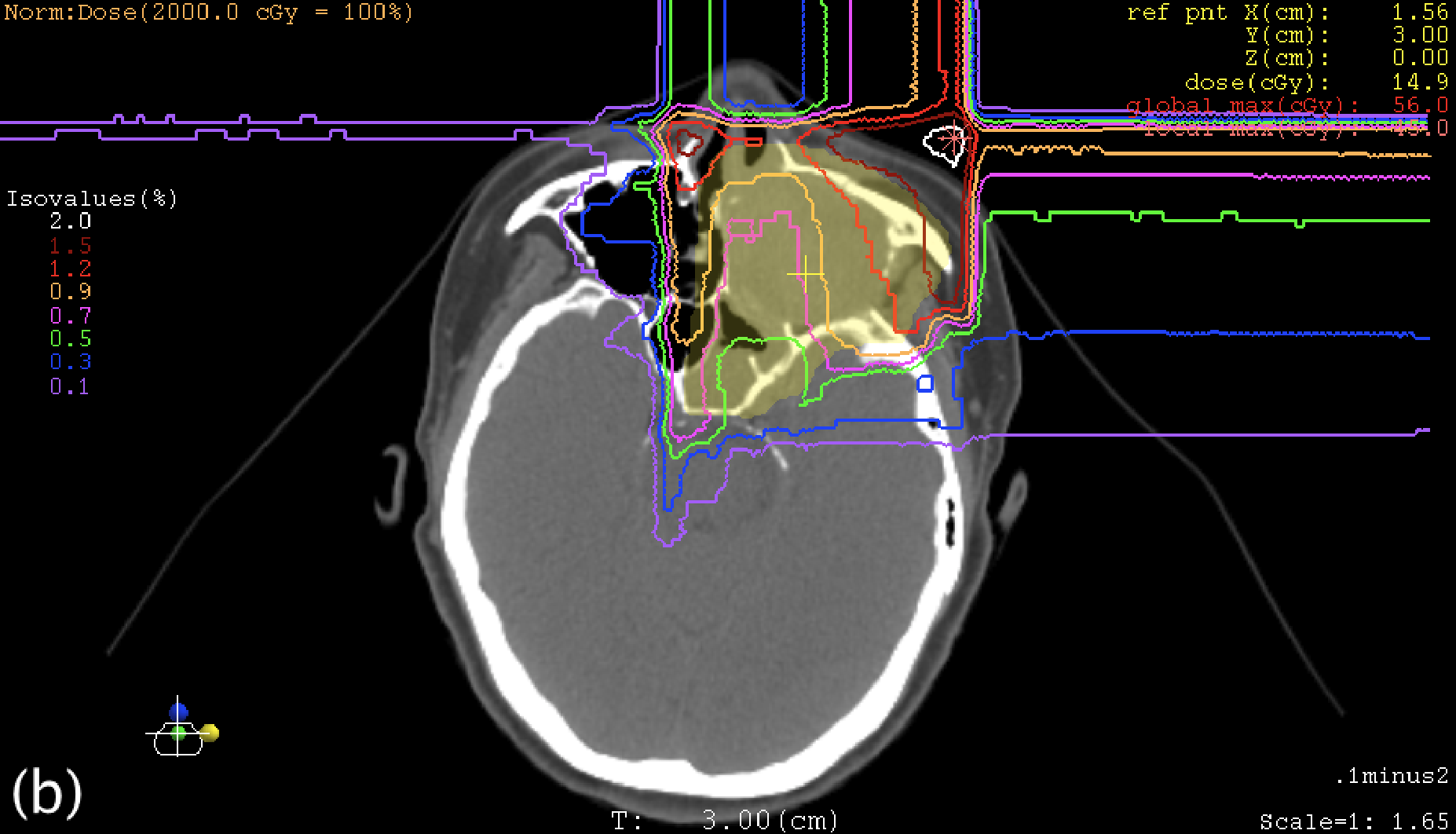}
\caption{\label{fig:5} Clinical-dose distribution on the isocenter ($+$) plane of trans-axial CT image for the nose-and-sinus target (yellowish region): (a) the modified-kernel plan with 105\%--10\% (white--purple) isodose contours and (b) dose reduction from the original-kernel plan with 2.0\%--0.1\% (white--purple) isodose contours.}
\end{figure}

Table~\ref{tab:3} shows the maximum WE thicknesses of the range compensators.
Except for the posterior beam in the sacrum case, range compensation was large as it had been expected for oblique-incident beams.
Figures \ref{fig:4} and \ref{fig:5} show the plan dose distributions, in which the reduction of clinical dose was 0.30\% at the isocenter in the sacrum case, 0.75\% at the isocenter in the nose-and-sinus case, and mostly within 1\% in these targets.
In the nose-and-sinus case, the dose reduction greater than 2\% occurred only in the air that happened to be included in the dose-calculation volume. 
Relatively speaking, the dose reduction was greater in the posterior part of the sacrum target and in the anterior-left (away from nose) part of the nose-and-sinus target, which was caused by locally shallower target depth and hence larger range compensation in the fields. 

\begin{table}
\caption{\label{tab:3} Maximum water-equivalent thickness $t_\mathrm{max}$ of range compensators designed for clinical cases.}
\setlength{\tabcolsep}{6pt}
\begin{tabular}{cllc}
\hline\hline
Case & Tumor site & Beam direction & $t_\mathrm{max}$/cm \\
\hline
(a) & Sacrum & Posterior & ~6.54 \\
 & & Right & 11.48 \\
 & & Left & 13.88 \\
 \hline
(b) & Nose and sinus & Anterior & ~9.40 \\
 & & Left & ~9.04 \\ 
\hline\hline
\end{tabular}
\end{table}

\section{Discussion}

Disagreement between the measured and calculated dose attenuation coefficients was found to be about 0.1\%/cm for the 290-MeV beams.
In the present model, the relative increase of the secondary-particle dose $D_2$ was equated with the relative decrease of the carbon-ion dose $D_1$ to correct energy dependence of the dose ratio $D_1/D_2$.
Although the inverse relation may be a natural assumption for the yield of secondary particles, their attenuation and variation of ionization with depth would disturb the assumed relation for dose.
In fact, the water thickness from the PE exit to the mid-SOBP depth is different between 290 MeV and 350 MeV by about 6 cm.
Due to reduced attenuation, the secondary-particle dose for 290 MeV could actually be larger than that expected from 350 MeV, which would have contributed to the observed disagreement.
Although a similar situation should apply inversely to the 400-MeV beams, no significant disagreement was found there somehow, which may indicate intrinsic limitations of the two-component model.
It was in principle possible to evaluate those parameters directly by redoing Monte Carlo simulation for ridge-filter design in accordance with the two-component model. 
However, it would be reasonable to assess the clinical influence of the PE effect that may possibly be overestimated by the simple calculation model.

In this study, we assumed invariant secondary-particle behavior between the mid-SOBP and tail-base depths, or $D_2 \approx D_\mathrm{tail}$ and $\epsilon_2 \approx \epsilon_\mathrm{tail}$, to deduce $D_1/D_2$ and $\epsilon_1/\epsilon_2$.
In a Monte Carlo study on a pristine 391-MeV carbon-ion beam in water,\cite{Kempe-2007} boron nuclei $^{10+11}$B showed notable dose contribution of about 1/4 of the tail-base dose, which would presumably be further moderated by range modulation.
If we assume $D_\mathrm{B} = 0.2\, D_\mathrm{tail} + \Delta D_\mathrm{B}$ for the mid-SOBP boron-nucleus dose with variation $\Delta D_\mathrm{B}$ among the range modulations and $\epsilon_\mathrm{B} \approx \epsilon_1 \approx 1.5 \, \epsilon_2$ for the boron-nucleus RBE, the dose ratio and the RBE ratio will vary relatively by 
\begin{eqnarray}
\frac{\Delta \frac{D_1}{D_2}}{\frac{D_1}{D_2}} = - \frac{\Delta D_2}{D_2} \approx -0.2\, \frac{\Delta D_\mathrm{B}}{D_\mathrm{B}} 
\end{eqnarray}
and
\begin{eqnarray}
\frac{\Delta \frac{\epsilon_1}{\epsilon_2}}{\frac{\epsilon_1}{\epsilon_2}} = - \frac{\Delta \epsilon_2}{\epsilon_2} = -\frac{(\epsilon_\mathrm{B}-\epsilon_2)\, \Delta D_\mathrm{B}}{D_2 + \Delta D_\mathrm{B}} \approx -0.1\, \frac{\Delta D_\mathrm{B}}{D_\mathrm{B}}. 
\end{eqnarray}
The reduced sensitivity to $\Delta D_\mathrm{B}$ supports the assumption of invariant secondary-particle behavior for their derivation. 
The Monte Carlo study also showed substantial dose from $^{10+11}$C nuclei.\cite{Kempe-2007}
Because all carbon nuclides are equivalent in ionization, the neutron-number variation resulted from neutron removal will only distort the Bragg peak, which will also be moderated by range modulation, and thus may be reasonably excluded from consideration.

Range compensators are usually concave-shaped with WE thickness less than a few centimeters in the middle of the field and large compensation only applies to the peripheral part.
In other words, reduction of clinical dose in the core part of a tumor will not be generally large.
For an ideal spherical target, the mean range compensation will be $\int_0^1 2 r \left(1- \sqrt{1-r^2} \right) dr = 1/3$ of the maximum, which may also be typical for irregular targets.
In an extreme case in which a beam is applied with range compensation of maximum 20 cm and of expected mean 6.8 cm, the PE effect on clinical dose could go up to 3\%.
In reality, however, the maximum range compensation is typically less than 10 cm and its effect on tumor dose will be further mitigated with multidirectional beams.
It is therefore reasonable that the PE effect on tumor dose in the clinical cases
resulted in about 1\% reduction.

In practical situations, where PE range compensators are always used, variation of the PE effect among patients may also be of the order of 1\%, which is normally smaller than dose fluctuation due to limitations of field uniformity or multiple scatting in heterogeneity.\cite{Yonai-2008b, Kanematsu-2011}
There are always uncertainties in treatment planning, patient setup, and beam delivery, which could cause dose errors typically up to a few percent.
In addition, the present radiobiology model is intrinsically limited in predicting clinical response in the presence of variability of radiosensitivity for dose fractionation, cell and tissue types, and individual patient, which could be even larger.
Nevertheless, the PE effect should be corrected in treatment planning rather than being handled as an uncertainty, or further the cause should be mitigated by using another material of better water equivalence such as polyoxymethylene.\cite{Kanematsu-2013}

\section{Conclusions}

We have investigated the effect of PE range compensators on tumor dose due to their non water equivalence in nuclear interactions.
In the model calculation, the extra carbon-ion attenuation per range shift of 0.45\%/cm caused reduction of 0.2\%--0.4\%/cm in RBE-weighted dose for clinical carbon-ion beams.
Reduction in absorbed dose was 0.1\%--0.3\%/cm, for which the measurements and calculations agreed with each other within 0.1\%/cm. 
The reduction of clinical dose due to PE range compensation could go up to 3\% in extreme cases. 
In a realistic treatment-planning study, it was typically within 1\% with expected variation of the same order.
That would be marginal as compared to intrinsic uncertainties originated in treatment planning, patient setup, beam delivery, and clinical response.

\begin{acknowledgments}
We thank the organizations and individuals that supported this study: Mitsubishi Electric Corporation for the framework of the treatment-planning system, K.~Yamamoto of Mitsubishi Space Software Co. Ltd. for software customization, W.~Furuichi of Accelerator Engineering Corporation for technical support in treatment planning, R.~Imai and A.~Hasegawa of the Radiation Oncology Section of NIRS for the clinical cases, and M.~Sekiguchi for scientific copyediting of the manuscript.

The experiment was carried out in the Research Project with Heavy Ions at NIRS-HIMAC 13H005 under principal investigator N.~Matsufuji.
\end{acknowledgments}


\begin{thebibliography}{99}

\bibitem{Urie-1983}
M.~Urie, M.~Goitein, and M.~Wagner, ``Compensating for inhomogeneities in proton radiation therapy,'' Phys. Med. Biol. {\bf 29}, 553--556 (1983). 

\bibitem{Kanai-1999}
T.~Kanai, M.~Endo, S.~Minohara, N.~Miyahara, H.~Koyama-Ito, H.~Tomura, N.~Matsufuji, Y.~Futami, A.~Fukumura, T.~Hiraoka, Y.~Furusawa, K.~Ando, M.~Suzuki, F.~Soga, and K.~Kawachi, ``Biophysical characteristics of HIMAC clinical irradiation system for heavy-ion radiation therapy,'' Int. J. Radiat. Oncol. Biol. Phys. {\bf 44}, 201--210 (1999).

\bibitem{Hishikawa-2004}
Y.~Hishikawa, Y.~Oda, H.~Mayahara, A.~Kawaguchi, K.~Kagawa, M.~Murakami, and M.~Abe, ``Status of the clinical work at Hyogo,'' Radiother. Oncol. {\bf 73}, S38--S40 (2004).

\bibitem{Ohno-2011}
T.~Ohno, T.~Kanai, S.~Yamada, K.~Yusa, M.~Tashiro, H.~Shimada, K.~Torikai, Y.~Yoshida, Y.~Kitada, H.~Katoh, T.~Ishii, and T.~Nakano, ``Carbon Ion Radiotherapy at the Gunma University Heavy Ion Medical Center: New Facility Set-up,'' Cancers {\bf 3}, 4046--4060 (2011).

\bibitem{Kanematsu-2013}
N.~Kanematsu, Y.~Koba, and R.~Ogata, ``Evaluation of plastic materials for range shifting, range compensation, and solid-phantom dosimetry in carbon-ion radiotherapy,'' Med. Phys. {\bf 40}, 041724-1--6 (2013).

\bibitem{Palmans-2002}
H.~Palmans, J.~ E.~Symons, J-M.~Denis, E.~A.~de~Kock, D.~T.~L.~Jones, and S.~Vynckier,``Fluence correction factors in plastic phantoms for clinical proton beams,'' Phys. Med. Biol. {\bf 47}, 3055--3071 (2002). 

\bibitem{Schneider-2002}
U.~Schneider, P.~Pemler, J.~Besserer, M.~Dellert, M.~Moosburger, J.~de~Boer, E.~Pedroni, and T.~Boehringer, ``The water equivalence of solid materials used for dosimetry with small proton beams,'' Med. Phys. {\bf 29}, 2946--2951 (2002).

\bibitem{Al-Sulaiti-2010}
L.~Al-Sulaiti, D.~Shipley, R.~Thomas, A.~Kacperek, P.~Regan, and H.~Palmans, ``Water equivalence of various materials for clinical proton dosimetry by experiment and Monte Carlo simulation,'' Nucl. Instrum. Meth. A {\bf 619}, 344--347 (2010). 

\bibitem{Luhr-2011}
A.~L\"uhr, D.~C.~Hansen, N.~Sobolevsky, H.~Palmans, S.~Rossomme, and N.~Bassler, ``Fluence correction factors and stopping power ratios for clinical ion beams,'' Acta Oncol. {\bf 50}, 797--805 (2011).

\bibitem{Moyers-2011}
M.~F.~Moyers, A.~S.~Vatnitsky, and S.~M.~Vatnitsky, ``Factors for converting dose measured in polystyrene phantoms to dose reported in water phantoms for incident proton beams,'' Med. Phys. {\bf 38}, 5799--5806 (2011).

\bibitem{Al-Sulaiti-2012}
L.~Al-Sulaiti, D.~Shipley, R.~Thomas, P.~Owen, A.~Kacperek, P.~Regan, and H.~Palmans, ``Water equivalence of some plastic-water phantom materials for clinical proton beam dosimetry,'' Appl. Radiat. Isotop. {\bf 70}, 1052--1055 (2012). 

\bibitem{Palmans-2013}
H.~Palmans, L.~Al-Sulaiti, P.~Andreo, D.~Shipley, A.~L\"uhr, N.~Bassler, J.~Martinkovi\v{c}, J.~Dobrovodsk\'y, S.~Rossomme, R.~A.~.S.~Thomas, and A.~Kacperek, ``Fluence correction factors for graphite calorimetry in a low-energy clinical proton beam: I. Analytical and Monte Carlo simulations,'' Phys. Med. Biol. {\bf 58}, 5363--5380 (2013).

\bibitem{Rossomme-2013}
S.~Rossomme, H.~Palmans, D.~Shipley, R.~Thomas, N.~Lee, F.~Romano, P.~Cirrone, G.~Cuttone, D.~Bertrand, and S.~Vynckier, ``Conversion from dose-to-graphite to dose-to-water in an 80 MeV/A carbon ion beam,'' Phys. Med. Biol. {\bf 58}, 5363--5380 (2013).

\bibitem{Torikoshi-2007} 
M.~Torikoshi, S.~Minohara, N.~Kanematsu, M.~Komori, M.~Kanazawa, K.~Noda, N.~Miyahara, H.~Ito, M.~Endo, and T.~Kanai, ``Irradiation System for HIMAC,'' J. Radiat. Res. (Tokyo) {\bf 48} (Suppl. A), A15--A25 (2007). 

\bibitem{Yonai-2008b}
S.~Yonai, N.~Kanematsu, M.~Komori, T.~Kanai, Y.~Takei, O.~Takahashi, Y.~Isobe, M.~Tashiro, H.~Koikegami, and H.~Tomita, ``Evaluation of beam wobbling methods for heavy-ion radiotherapy,'' Med. Phys. {\bf 35}, 927--938 (2008).

\bibitem{Kase-2006}
Y.~Kase, N.~Kanematsu, T.~Kanai, and N.~Matsufuji, ``Biological dose calculation with Monte Carlo physics simulation for heavy-ion radiotherapy,'' Phys. Med. Biol. {\bf 51}, N467--N475 (2006).

\bibitem{Sakama-2012}
M.~Sakama, T.~Kanai, Y.~Kase, K.~Yusa, M.~Tashiro, K.~Torikai, H.~Shimada, S.~Yamada, T.~Ohno, and T.~Nakano, ``Design of ridge filters for spread-out Bragg peaks with Monte Carlo simulation in carbon ion therapy,'' Phys. Med. Biol. {bf 57}, 6615--6633 (2012).

\bibitem{Kanematsu-2007}
N.~Kanematsu, M.~Torikoshi, M.~Mizota, and T.~Kanai, ``Secondary range shifting with range compensator for reduction of beam data library in heavy-ion radiotherapy,'' Med. Phys. {\bf 34}, 1907--1910 (2007).

\bibitem{Bortfeld-1997}
T.~Bortfeld, ``An analytial approximation of the Bragg curve for therapeutic proton beams,'' Med. Phys. {\bf 24}, 2024--2033 (1997).

\bibitem{Matsufuji-2003}
N.~Matsufuji, A.~Fukumura, M.~Komori, T.~Kanai, and T.~Kohno, ``Influence of fragment reaction of relativistic heavy charged particles on heavy-ion radiotherapy,'' Phys. Med. Biol. {\bf 48}, 1605--1623 (2003).

\bibitem{Goldhaber-1974}
A.~S.~Goldhaber, ``Statistical models of fragmentation processes,'' Phys. Lett. B {\bf 53}, 306--308 (1974).
 
\bibitem{Matsufuji-2005}
N.~Matsufuji, M.~Komori, H.~Sasaki, K.~Akifu, M.~Ogawa, A.~Fukumura, E.~Urakabe, T.~Inaniwa, T.~Nishio, T.~Kohno, and T.~Kanai, ``Spatial fragment distribution from a therapeutic pencil-like carbon beam in water,'' Phys. Med. Biol. {\bf 50}, 3393--3403 (2005).

\bibitem{Yonai-2008a}
S.~Yonai, N.~Matsufuji, T.~Kanai, Y.~Matsui, K.~Matsushita, H.~Yamashita, M.~Numano, T.~Sakae, T.~Terunuma, T.~Nishio, R.~Kohno, and T.~Akagi, ``Measurement of neutron ambient dose equivalent in passive carbon-ion and proton radiotherapies,'' Med. Phys. {\bf 35}, 4782--4792 (2008).

\bibitem{Furusawa-2000}
Y.~Furusawa, K.~Fukutsu, M.~Aoki, H.~Itsukaichi, K.~Eguchi-Kasai, H.~Ohara, F.~Yatagai, T.~Kanai, and K.~Ando, ``Inactivation of aerobic and hypoxic cells from three different cell lines by accelerated 3He-, 12C- and 20Ne-ion beams,'' Radiat. Res. {\bf 154}, 485-496 (2000).

\bibitem{Lam-1987}
G.~K.~Y.~Lam, ``The survival response of a biological system to mixed radiations,'' Radiat. Res. {\bf 110}, 232--243 (1987).

\bibitem{Tilly-1999}
N.~Tilly, A.~Brahme, J.~Carlsson, B.~Glimelius, ``Comparison of cell survival models for mixed LET radiation,'' Int. J. Radiat. Biol. {\bf 75}, 233--243 (1999).

\bibitem{Matsufuji-2007}
N.~Matsufuji, T.~Kanai, N.~Kanematsu, T.~Miyamoto, M.~Baba, T.~Kamada, H.~Kato, S.~Yamada, J.~Mizoe, and H.~Tsujii, ``Specification of carbon ion dose at the National Institute of Radiological Sciences (NIRS),'' J. Radiat. Res. {\bf 48} (Suppl. A), A81--A86 (2007).

\bibitem{Kanematsu-2008}
N.~Kanematsu, ``Alternative scattering power for Gaussian beam model of heavy charged particles,'' Nucl. Instrum. Methods B {\bf 266}, 5056--5062 (2008).

\bibitem{Petti-1992}
P.~L.~Petti, ``Differential-pencil-beam dose calculations for charged particles,'' Med. Phys. {\bf 19}, 137--149 (1992).

\bibitem{Tomura-1998}
H.~Tomura, T.~Kanai, A.~Higashi, Y.~Futami, N.~Matsufuji, M.~Endo, F.~Soga, and K.~Kawachi, ``Analysis of the penumbra for uniform irradiation fields delivered by a wobbler method,'' Japan. J. Med. Phys. {\bf 18}, 42--56 (1998).

\bibitem{Kanematsu-2006}
N.~Kanematsu, T.~Akagi, Y.~Takatani, S.~Yonai, H.~Sakamoto, H.~Yamashita, 
``Extended collimator model for pencil-beam dose calculation in proton radiotherapy,'' Phys. Med. Biol. {\bf 51}, 4807--4817 (2006).

\bibitem{Kanematsu-2009}
N.~Kanematsu, ``Semi-empirical formulation of multiple scattering for the Gaussian beam model of heavy charged particles stopping in tissue-like matter,'' Phys. Med. Biol. {\bf 54}, N67--N73 (2009).

\bibitem{Kempe-2007}
J.~Kempe, I.~Gudowska, and A.~Brahme, ``Depth absorbed dose and LET distributions of therapeutic $^1$H, $^4$He, $^7$Li, and $^{12}$C beams,'' Med. Phys. {\bf 34}, 183--192 (2007).

\bibitem{Kanematsu-2011}
N.~Kanematsu, ``Modeling of beam customization devices in the pencil-beam splitting algorithm for heavy charged particle radiotherapy,'' Phys. Med. Biol. {\bf 56}, 1361--1371 (2011).

\end{thebibliography}
\end{document}